\documentclass[draftclsnofoot, onecolumn, comsoc, 12pt]{IEEEtran} 

\usepackage{amsmath,amssymb,amsfonts}

\usepackage{caption}
\usepackage{subcaption}

\usepackage{stfloats}
\usepackage{float}
\usepackage{wrapfig}
\usepackage{xcolor}
\usepackage{colortbl} 
\usepackage{color,soul}
\usepackage{url}
\usepackage{verbatim}
\usepackage{graphicx}
\usepackage{cite}

\usepackage{multirow}
\usepackage{array,boldline, makecell,booktabs}
\usepackage{longtable} 

    \usepackage[group-separator={,},group-minimum-digits=4]{siunitx}

\begin{document}
%
\title{Synergizing Airborne Non-Terrestrial Networks and Reconfigurable Intelligent Surfaces-Aided 6G IoT}
%
%
%

\author{Muhammad Ali Jamshed, Aryan Kaushik, Mesut Toka, Wonjae Shin, \\Muhammad Zeeshan Shakir, Soumya P. Dash, and Davide Dardari
\thanks{
M. A. Jamshed is with the College of Science and Engineering, University	of Glasgow, UK (e-mail: \{muhammadali.jamshed@glasgow.ac.uk).\\
$~~~$A. Kaushik is with the School of Engineering \& Informatics, University	of Sussex, UK (e-mail: aryan.kaushik@sussex.ac.uk). \\
$~~~$M. Toka is with the Department of Electrical and Computer Engineering, Ajou University, South Korea (e-mail:tokamesut@ajou.ac.kr).\\
$~~~$W. Shin is with the School of Electrical Engineering, Korea University, South Korea, and the Department of Electrical and Computer Engineering, Princeton University, USA (e-mail: wjshin@korea.ac.kr).\\
$~~~$M. Z. Shakir is with the School of Computing, Engineering and Physical Sciences, University of the West of Scotland, UK (e-mail: \{muhammad.shakir@uws.ac.uk).\\
$~~~$S. P. Dash is with the School of Electrical Sciences, Indian Institute of Technology Bhubaneswar, India (e-mail: spdash@iitbbs.ac.in). \\
$~~~$D. Dardari is with the Department of Electrical, Electronic, and Information Engineering, University of Bologna, Italy (e-mail: davide.dardari@unibo.it). \\
}}

%
%


\maketitle
\vspace{-5mm}
\begin{abstract}
On the one hand, Reconfigurable Intelligent Surfaces (RISs) emerge as a promising solution to meet the demand for higher data rates, improved coverage, and efficient spectrum utilization. On the other hand, Non-Terrestrial Networks (NTNs) offer unprecedented possibilities for global connectivity. Moreover, the NTN can also support the upsurge in the number of Internet of Things (IoT) devices by providing reliable and ubiquitous connectivity. Although NTNs have shown promising results, there are several challenges associated with their usage, such as signal propagation delays, interference, security, etc. In this article, we have discussed the possibilities of integrating RIS with an NTN platform to overcome the issues associated with NTN. Furthermore, through experimental validation, we have demonstrated that the RIS-assisted NTN can play a pivotal role in improving the performance of the entire communication system. 

\end{abstract}

 \begin{IEEEkeywords}
Non-Terrestrial Networks (NTNs), Unmanned Aerial Vehicles (UAVs), Reconfigurable Intelligent Surfaces (RIS), Internet of Things (IoT).
 \end{IEEEkeywords}

\IEEEpeerreviewmaketitle

\vspace{-3mm}

\section{Introduction}
\label{sec:1}
The rapid evolution of wireless communication technologies has profoundly shaped the way we interact, work, and access information. From the early days of voice-only communication to the era of high-speed data transfer and Internet of Things (IoT) connectivity, wireless communication has become an indispensable part of modern life \cite{MojtabaIoT}. However, as the demand for higher data rates, improved coverage, and efficient spectrum utilization continues to grow, traditional communication infrastructure faces limitations. Reconfigurable Intelligent Surfaces (RIS), a groundbreaking concept, emerges as a promising solution to overcome these challenges and reshape the landscape of wireless communications.

RIS, also known as smart surfaces, encompasses a new paradigm in wireless communication technology. Unlike traditional antennas and base stations (BSs), which transmit and receive signals directly, RIS operates as passive elements that reflect, refract, and manipulate electromagnetic waves. These surfaces are equipped with an array of controllable elements that, when dynamically controlled, empowers RIS to modify signal propagation characteristics, leading to improved signal strength, coverage extension, interference mitigation, and energy efficiency. This has resulted in several applications of RIS-aided communications in 5G and beyond technologies, IoT, wireless virtual and augmented reality, connected vehicles, and disaster recovery and remote areas coverage \cite{WeiShiIoE}.

Another significant development in recent years for next-generation communication systems is the concept of Non-Terrestrial Networks (NTNs), which heralds a new era of connectivity by transcending the confines of Earth's surface \cite{Gue:20}. NTNs encompass an array of innovative communication systems that operate beyond traditional terrestrial infrastructures, offering unprecedented possibilities for global connectivity via the internet, IoT, navigation, disaster resilience, remote access, Earth observation, and other scientific explorations such as interplanetary communications.

NTNs are typically deployed in space or the upper atmosphere, utilizing satellites, high-altitude platforms (HAPs), or other spaceborne assets. Typically, low Earth orbit (LEO), medium Earth orbit (MEO), and geostationary orbit (GEO) satellites are deployed for NTN systems. LEO satellites orbit at altitudes ranging from a few hundred kilometers to around 2,000 kilometers. They offer low latency and high-speed connectivity, making them ideal for services such as global internet coverage, remote sensing, and real-time data transmission. GEO satellites, positioned at an altitude of approximately 35,786 kilometers, remain stationary relative to a specific point on Earth's surface and are commonly used for broadcasting, communication, and weather observation due to their stable coverage area. Further, MEO satellites, operating at altitudes between LEO and GEO orbits provide a balance between coverage and latency. Global navigation systems like GPS and Galileo leverage MEO satellites for accurate positioning and timing services. Apart from satellites, NTNs use HAPs which are vehicles stationed in the stratosphere at altitudes between 20 to 50 kilometers. HAPs can serve as relays for communication between ground stations or provide wireless connectivity to remote or disaster-stricken areas. They offer advantages such as longer mission duration compared to satellites and the ability to cover specific regions efficiently.

Although NTNs can support the expansion of the IoT by providing reliable and ubiquitous connectivity for countless interconnected devices, there are several challenges associated with their usage. The growing number of satellites raises concerns about orbital debris and effective spectrum allocation to prevent interference. While LEO satellites offer low latency, GEO satellites suffer from signal propagation delays, and thus, interference management becomes a crucial challenge to overcome to ensure reliable and uninterrupted connectivity. However, combining the capabilities of RIS and NTNs can lead to superior coverage and signal quality. Strategic positioning of RIS in areas with coverage gaps would result in reflecting and redirecting signals from NTNs to areas that were previously underserved. This integration can also effectively extend the reach of NTNs to urban canyons, dense forests, and remote regions, bridging the digital divide and providing consistent connectivity to a wider range of locations. Furthermore, by deploying RIS in conjunction with NTNs, interference from surrounding environments can be mitigated through precise signal redirection, which would result in higher data rates and improved signal-to-noise (SNR) ratios, leading to a more reliable and seamless user experience in IoT networks. Moreover, the dynamic nature of both RIS and NTNs enables real-time adaptation to changing environmental conditions.

\section{RIS-aided NTN IoT for 6G and Beyond: Advances and Capabilities}
\label{sec:2}
By considering the IoT-dominated communication networks in the future, which is also one of the 6G visions, the use of satellites (LEO, MEO, and GEO) is undoubtedly required. This is because billions of cellular IoT connections including all the types such as massive, critical, industrial, and broadband are expected \cite{MojtabaIoT}. On the other hand, the RIS has remarkable features in that it can enhance wireless communication in urban and industrial areas, combating multi-path fading, attenuation, and interference, and support ubiquitous IoT connectivity under Industry 4.0 and 5.0 envisioned by 6G. With the noticeable recent developments, the most promising capabilities/advances of RIS-aided satellite communications can be highlighted below: 

\textbf{Reduced Complexity via Over-the-Air Processing:} In the case of LEO-aided communication, the ground receivers should have steering antennas with beam tracking hardware due to the high velocity of LEOs. Unfortunately, IoT devices cannot handle such operations that require highly complex signal processing due to their limited and energy-efficient transceiver structures. To deal with this issue, the deployment of an RIS on the satellite can be an appropriate solution, which yields eliminating possible extra signal processing burden on IoT devices owing to over-the-air processing \cite{WeiShiIoE}. On the other hand, the deployment of the RIS on any building also exhibits the same benefits.  

\textbf{Adaptive Coverage:} Recent experimental developments on the adjustment of RIS phases, as conducted in \cite{TangWRISPrac}, have shed light on enabling flexible and adaptive coverage with respect to dense IoT networks. This is due to the reconfigurable features of the RIS elements. Therefore, with the deployment of the RIS on the satellite, the service coverage can be modified in an adaptive manner by the means of beam directing. By doing so, not only RIS-aided broadcasting but also serving to specific IoT devices requiring high priority can be possible.

\textbf{High Spatial Resolution:} RISs are also capable of providing high spatial resolution. Specifically, in addition to improving the quality of communication, RISs can take a role to eliminate the overhead of sensing/localization. For instance, RIS-mounted LEOs can satisfy the communication/sensing demands of emergency incidents in remote areas, forests, deserts, and seas. Nevertheless, it becomes cumbersome when space-air-ground-sea integrated networks including densely IoT applications are considered in terms of a vast spherical area. This is because RISs have limitations in sensing/localization. Therefore, most recently, holographic RISs have attracted much attention \cite{HoloRIS}, to address these limitations. 

\textbf{Enhanced Secure Information-Exchange:} 
With the ability to control electromagnetic waves, RISs can enhance the security among IoT devices in the physical layer medium. To do this, they can modify the phase shifts to maximize the SNR ratio received by the legitimate user. This can be also possible if they are deployed at both satellites and terrestrial structures. More importantly, RIS-mounted satellites can act as a jammer incorporating with the ground BS and send artificial noise signals through the illegitimate users \cite{WeiShiIoE}. 

\section{RIS-Aided Airborne NTNs: Use Cases and Ubiquitous Connectivity}
\label{sec:3}

\subsection{RIS-Aided Satellite NTNs and Interference Management}
\label{sec:3.1}

Thanks to the recent advances in academia and industry, RIS technology has become an indispensable concept for the vision of 5G and beyond. Additionally, considering the limitations of the terrestrial infrastructures, it has been proven by academia and commercial applications that satellites play a crucial role in 5G and beyond communications, not only in rural but also in urban areas. Therefore, the combination of the RIS technology with satellite NTNs has become a paramount focus to combat backhauling challenges resulting from ultra-massive IoT networks.

\begin{figure}[t!]
    \begin{center}    \vspace{-2mm}
        \includegraphics[width=\columnwidth]{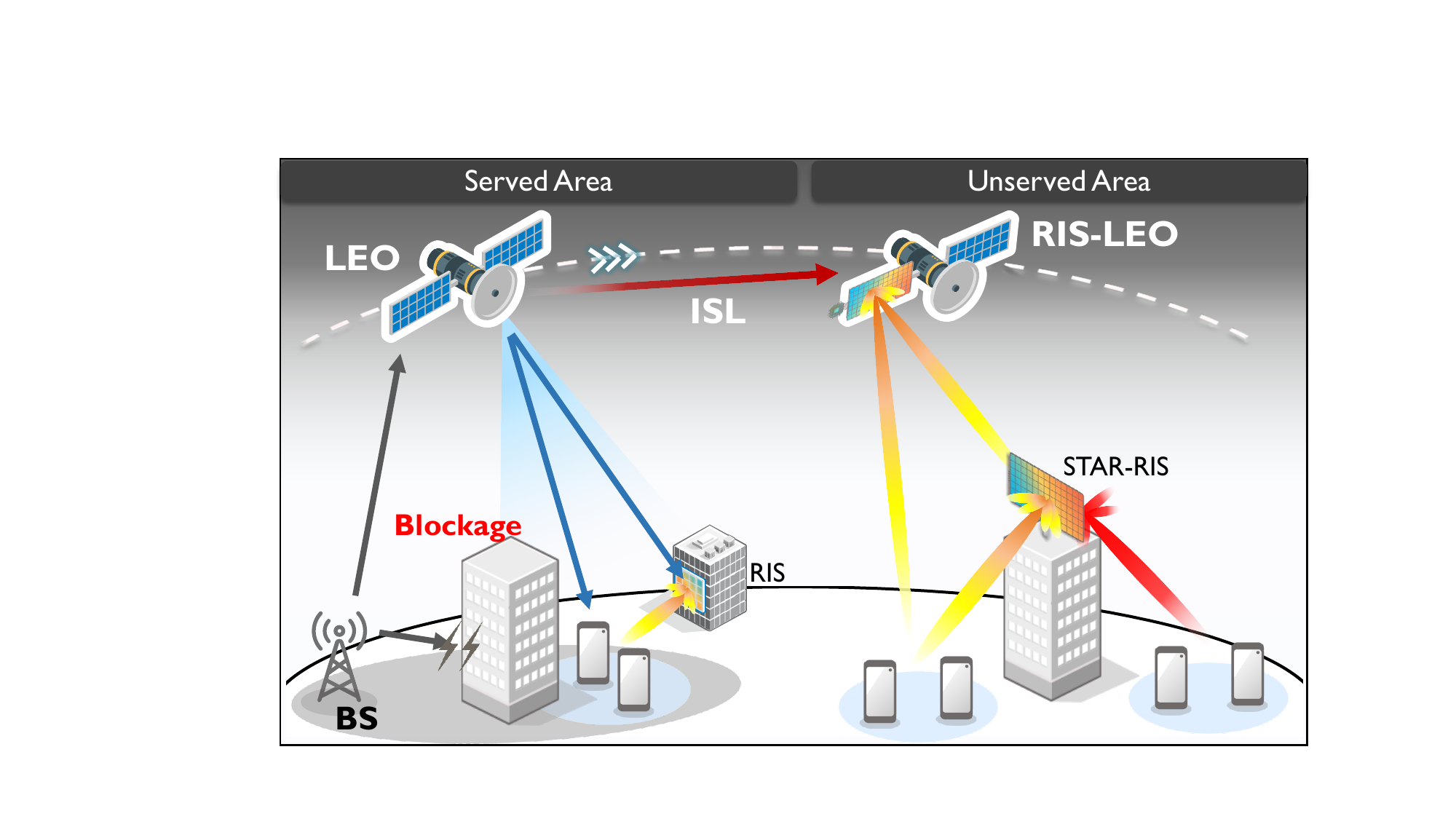}
    \end{center}
    \setlength{\belowcaptionskip}{-2pt}\vspace{-1mm}
    \caption{RIS-aided satellite NTN.}
    \vspace{-4mm}
    \label{fig1RISLEO}
\end{figure}

According to the demanded usage purposes, there are many deployment scenarios of RISs in satellite-based NTNs. First of all, in all deployment scenarios, the RIS provides a common benefit which is link quality enhancement. Because, with many RIS elements, the receiver can receive many induced-modified replicas of the transmitted information. For instance, as shown in Fig. \ref{fig1RISLEO}, link quality can be poor between the satellite and ground users; thus, a terrestrial RIS can assist the satellite in improving the quality of the received information by the ground users. To make ubiquitous connectivity possible not only in served but also in unserved areas, one of the promising approaches is to exploit RIS-deployed LEOs (as shown in Fig. \ref{fig1RISLEO}). Thereby, the LEO within the served area can transmit the corresponding information to the RIS-deployed LEO  to serve IoT users in unserved areas. This can be possible via an inter-satellite link (LEO) between LEOs. This approach reduces the number of unnecessary costly terrestrial infrastructures thanks to the nearby RIS-LEO satellites. Given the recent commercial attempts to launch many LEO satellites, it is reasonable to deploy RIS units on the satellites that move at the same or different orbits. Additionally, terrestrial RIS nodes, including simultaneous transmission and reflection (STAR)-RISs can be deployed to further increase the coverage and link improvement. Another paramount use case is to exploit RIS nodes in a spectrum-sharing integrated satellite-terrestrial network wherein the satellites and terrestrial infrastructures have the opportunity to use the same spectrum resources. However, interference exists from satellite links through the RIS node in the terrestrial links. Moreover, if a multi-beam scenario is considered, the ground users also suffer from inter-/intra-beam interference severely. Fortunately, the RIS allows interference signals to be controlled physically; thus, the interference can be nullified for the corresponding beams by appropriately adjusting the phases. Nevertheless, this remains a challenging issue because signal processing overhead can increase, and appropriate phase adjustment may not always be possible.               

\subsection{RIS-Aided UAVs}
\label{sec:3.2}

Unmanned aerial vehicles (UAVs) offer intriguing possibilities to enhance the flexibility and scalability of future networks. UAVs possess the potential to fly on-demand and precisely where needed while also being capable of responding swiftly to fluctuations in traffic demands, relying on a dominant line-of-sight (LoS). One of the most extensively explored applications of UAVs is their integration as mobile BSs, functioning as complementary aerial platforms within the low-altitude airspace (below 150 m) of 5G/6G cellular networks to bolster communication services or establish a dynamic radar network \cite{Gue:20}. This utilization of UAVs as mobile BSs is particularly well-suited for serving massive machine-type communication (mMTC) and IoT links, given that IoT nodes are primarily stationary (with fixed positions over time) and typically exhibit predictable traffic demands \cite{Mig:21}. Additionally, deploying UAVs in closer proximity to these nodes ensures reduced latency. However, it is important to note that this solution does have certain limitations, primarily related to the high energy consumption and weight of the BSs, which may not align with the imperative for extended battery life and compact size of the UAVs.

\begin{figure}[t!]
    \begin{center}    
        \includegraphics[width=\columnwidth]{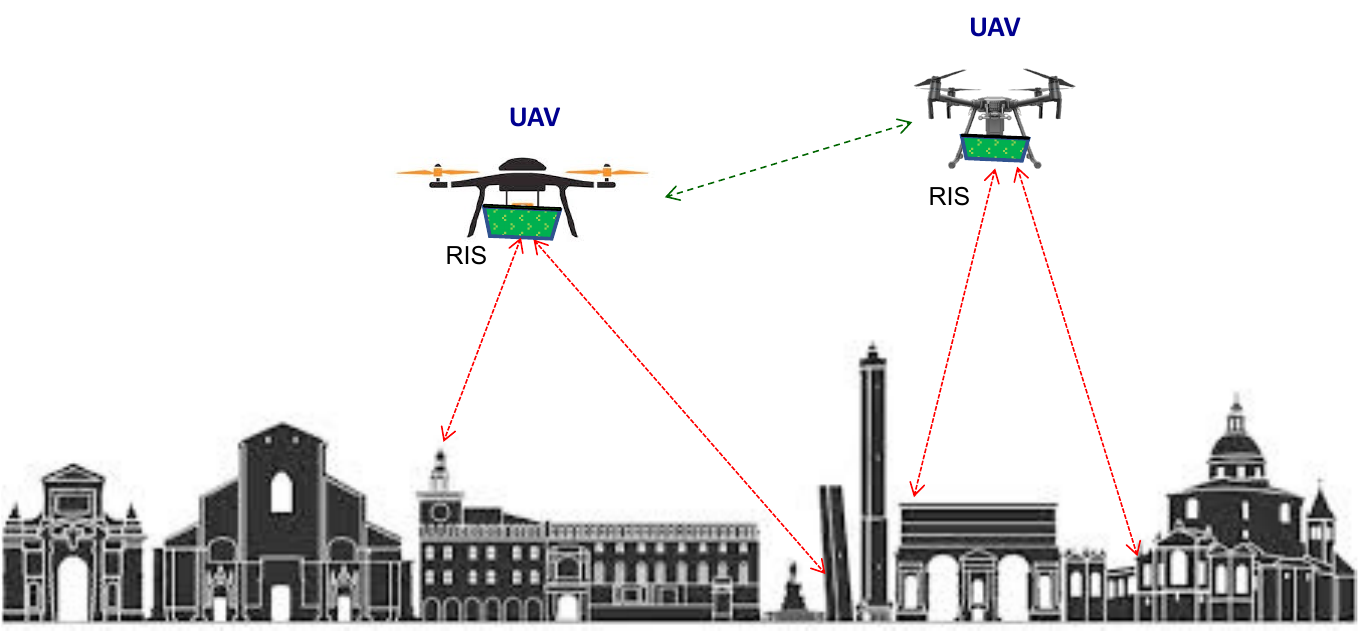}
    \end{center}
    \setlength{\belowcaptionskip}{-2pt}\vspace{-1mm}
    \caption{RIS-aided UAV communications.}
    \label{figUAVRIS}
\end{figure}

Recently, the possibility of mounting a RIS on UAVs to function as passive smart reflectors has been put forth in the literature \cite{Sha:21,Liu:22}, see Fig. \ref{figUAVRIS}. This approach offers several key advantages, including reduced weight, lower costs, negligible energy consumption, and efficient spectrum utilization (as it does not involve emitting additional radio frequency (RF) waves). Consequently, it addresses many of the previously mentioned limitations associated with flying BSs. Furthermore, RISs possess inherent full-duplex capabilities and do not introduce latency into communication processes. However, several challenges must be addressed before UAV-RIS-aided wireless networks can become a viable solution. Below, we summarize the main issues and areas of investigation: 

\textbf{Poor Link Budget:} As non-regenerative relays, UAV-RIS networks experience total path loss determined by the cascade of forward and backward links. To mitigate this substantial path loss, large RISs with precise beamforming capabilities are required. Recently, active RISs have emerged as potential solutions, but they may draw additional energy from UAV batteries.

\textbf{Fast Channel Estimation and Tracking:} The movement of UAVs and the vibrations that generate micro-Doppler effects make high-frequency channel estimation and precise beamforming extremely challenging. Considering that RISs lack sensors and channel estimation must occur at the cascaded channel sides, this poses a significant challenge. Furthermore, based on the channel estimate, the RIS must be configured at high speeds, potentially increasing hardware complexity and signaling overhead. Statistical-based approaches that optimize and configure the RIS based on long-term channel measurements accounting for UAV and IoT device mobility models, show promise in addressing this issue \cite{AbrDarDiR:J21}.

\textbf{Anomalous Reflections, Interference, and Jamming:} When the RIS' phase profile is set to reflect a signal (or multiple signals) from a known direction (e.g., from the BS) to another direction (e.g., toward the IoT device), it cannot prevent the reflection of signals from other unknown RF sources in unintended directions, leading to interference. This characteristic makes UAV-RIS-aided wireless networks vulnerable to jamming if not properly counteracted \cite{Tan:21}. Introducing non-diagonal RISs, where the coupling between elements can also be configured, potentially yielding a non-diagonal transfer matrix, may offer a solution to prevent the RIS from reflecting signals in unintended directions.

\textbf{Path Optimization:} UAV path optimization is a distinctive aspect of UAV-based communications and has been the subject of numerous research papers \cite{Liu:22}. If a global UAV-RIS wireless network is to be achieved, path optimization must be jointly addressed alongside the aforementioned challenges.

\begin{figure*}[t!]
\centering
\includegraphics[width=0.85\textwidth]{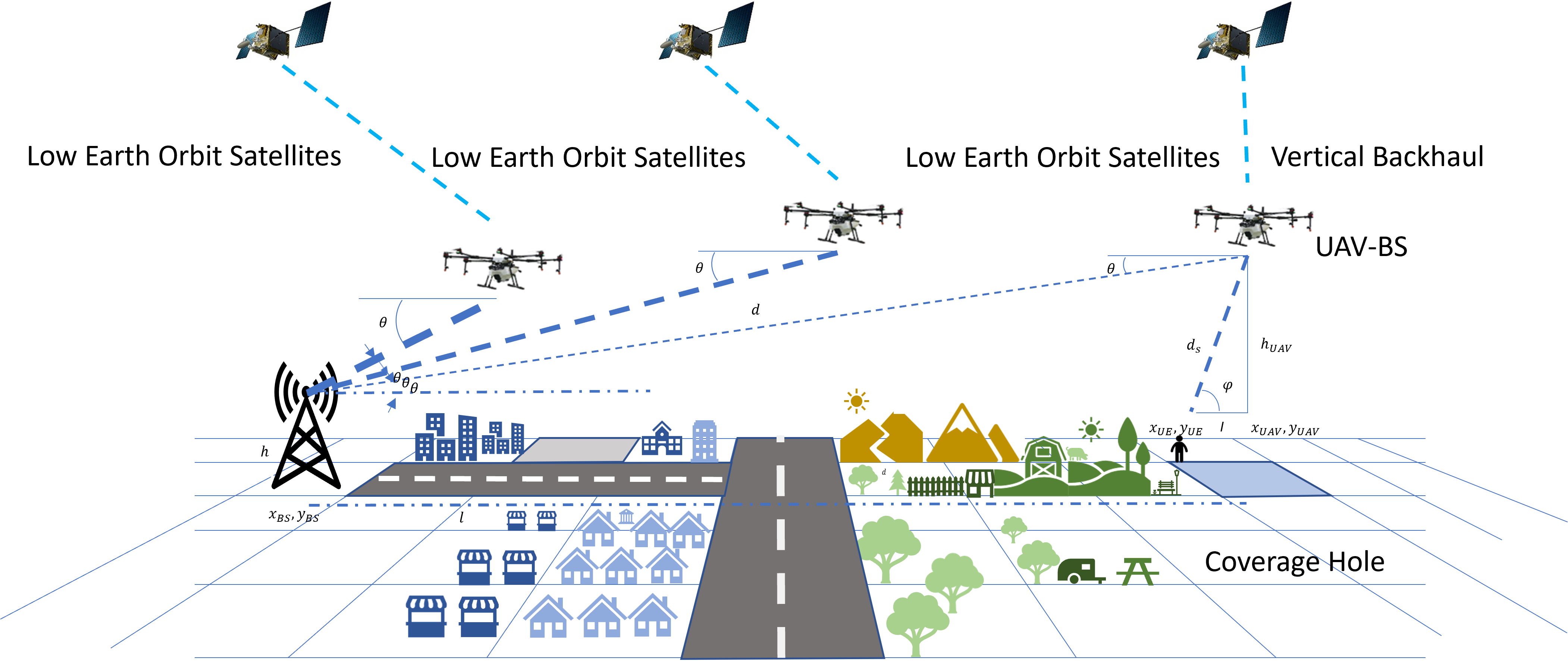}
    \caption{Graphical illustration of the RIS-assisted UAV network for coverage hole discovery and network capacity expansion in both planned and emergency scenarios.}
    \label{fig3a}
\end{figure*}

\subsection{RIS and AI-aided UAVs for Ubiquitous Connectivity and Coverage Hole Discovery}
\label{sec:3.3}
Two comprehensive scenarios illustrate the integration of flying platforms into future networks, with RIS and machine learning algorithms assuming a pivotal role in network densification:
\begin{itemize}

   \item Deployment Configuration 1 (UAV as a Coverage Hole detector): In this scenario, RIS-assisted flying platforms are poised to complement traditional cellular networks, augmenting wireless capacity, expanding coverage, and enhancing network reliability, particularly during temporary events, typically in remote, hard-to-reach regions where conventional infrastructure is scarce and prohibitively expensive to establish. Intelligent algorithms come into play by operating in tandem with flying platforms to identify and anticipate network outages, facilitating the swift deployment of flying BSs. For instance, our previous work \cite{UAVcoveragehole2020} leveraged reinforcement learning to achieve model-free learning about outages across diverse radio environments. This approach enabled us to explore the radio environment effectively, predict coverage gaps, and deploy UAVs connected to terrestrial BSs and LEO satellites, as depicted in Fig. \ref{fig3a}. 

   \item Deployment Configuration 2 (UAV as an Access Network): In this alternative scenario, RIS-assisted flying platforms are mobilized to provide access network as a flying small cell or a hub to aggregate the traffic from small cells to the core network for unforeseen emergencies, such as disaster relief operations or when conventional cellular networks suffer damage or congestion. Benefiting from their mobility, flying platforms can be rapidly and efficiently deployed to reinforce cellular networks, elevate network quality-of-service (QoS), and bolster network resilience in critical situations \cite{UAVasAccess}. Here, intelligent algorithms come into play once more, this time focusing on predicting optimal trajectories for flying base stations to establish pop-up networks swiftly. Techniques like deep learning prove valuable for obstacle detection, facilitating the rapid deployment of communication infrastructure in times of crisis. 
\end{itemize}

These two deployment configurations underscore the transformative potential of flying platforms in revolutionizing network capabilities, especially when synergized with machine learning algorithms. RIS-assisted wireless communication between LEO satellites and flying platforms addresses the evolving demands of modern communication networks, serving both planned and emergency scenarios with resilience and efficiency.

\section{RIS-UAV IoT  Design/Results}
\label{sec:4}

To validate the significance of integrating RIS with the NTN platform, in this section, we have considered a downlink scenario, where a BS establishes a LoS communication link with an IoT device by utilizing a UAV equipped with RIS. In our experimental analysis, we assess the achievable data rate performance utilizing the semi-definite relaxation (SDR) approach and juxtapose it against the scenario where the RIS surface remains unconfigured. The simulation settings are aligned with the 3GPP channel model \cite{3gpp1}. We examine an orthogonal frequency division multiplexing (OFDM) system featuring 1000 subcarriers and a pragmatic multi-path channel model characterized by 23 channel taps at random delays with an exponential power decay profile. The locations of the BS, UAV, and IoT devices are [20 -300 0], [0 0 100], and [20 0 0], respectively. Our focus involves LoS channels between the BS-UAV connections. This selection ensures reliable data transmissions and substantial channel gains. Conversely, we presume that a non-LoS (NLoS) propagation takes place between the BS and the IoT device.

\begin{figure}
     \centering
     \begin{subfigure}[b]{0.45\textwidth}
         \centering
         \includegraphics[width=\columnwidth]{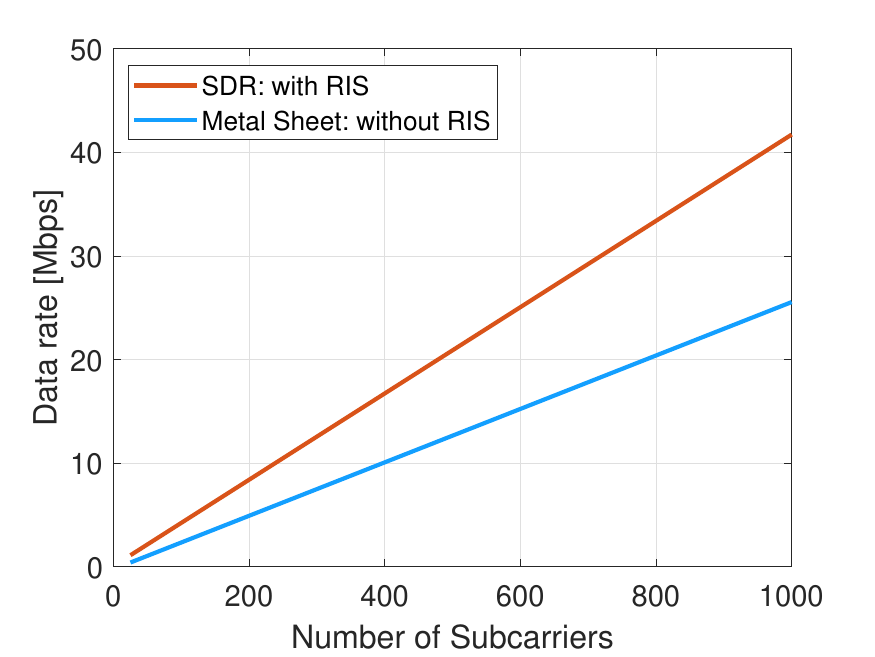}
         \caption{Comparative analysis of achievable data rate [Mbps] versus number of subcarriers.}
         \label{fig1}
     \end{subfigure}
     \vfill
     \begin{subfigure}[b]{0.45\textwidth}
         \centering
         \includegraphics[width=\columnwidth]{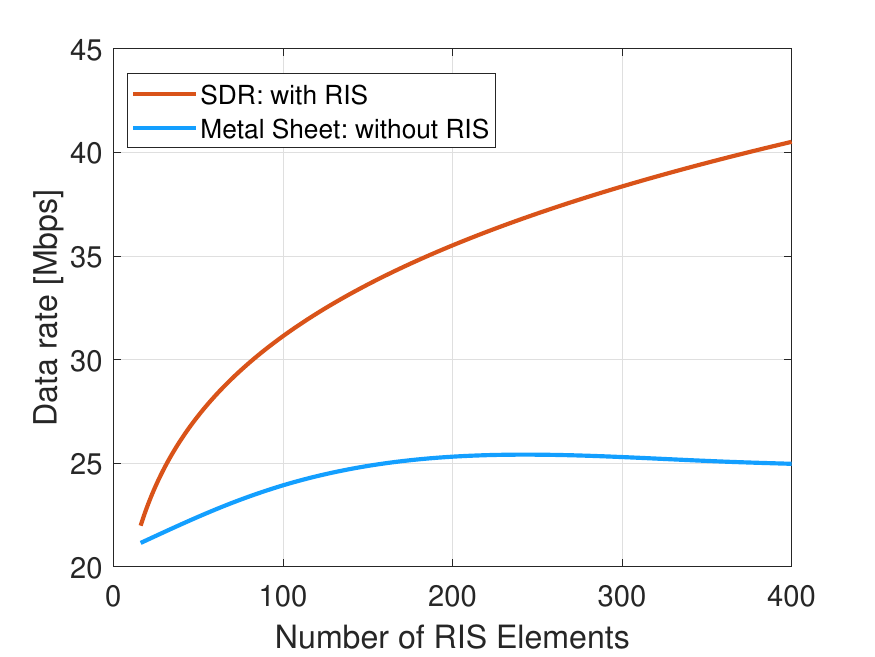}
         \caption{Comparative analysis of achievable data rate [Mbps] versus number of RIS elements.}
         \label{fig2}
     \end{subfigure}
     \vfill
     \begin{subfigure}[b]{0.45\textwidth}
         \centering
         \includegraphics[width=\columnwidth]{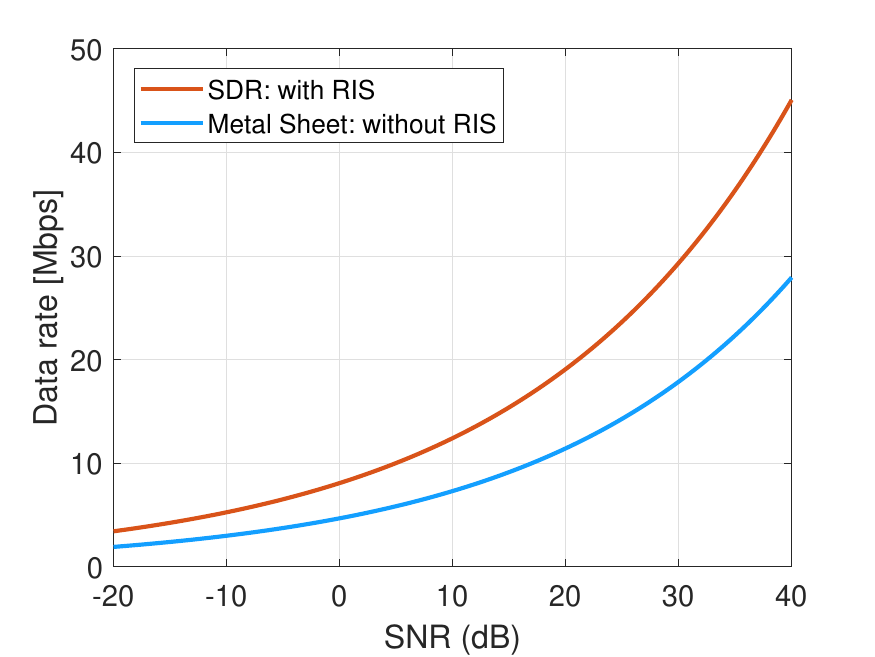}
         \caption{Comparative analysis of achievable data rate [Mbps] versus SNR.}
         \label{fig3}
     \end{subfigure}
        \caption{Comparative analysis of RIS-aided UAV for cases: (1) when RIS based SDR method is employed, and (2) when the surface is unconfigured (passive metal sheet).}
\end{figure}

In Fig. \ref{fig1}, we compare the achievable data rate of the SDR method with the unconfigured RIS for a fixed number of RIS elements, i.e., 400 and varying the number of subcarriers. The SDR method outperforms the unconfigured surface. In both cases, the achievable data rate is dramatically boosted as the range of subcarriers increases. In comparison to unconfigured RIS, the utilization of the SDR method, along with a fined-tuned RIS configuration, leads to a remarkable enhancement in the total data rate. This improvement is achieved by optimizing the coefficients of the RIS elements. These optimized coefficients enable more efficient and synergistic integration of both the direct communication path and the reflected path via the RIS at the receiver. As a result, the overall data transmission rate is substantially increased.

Figure \ref{fig2} presents the achievable data rate against the number of RIS elements for the scenarios when the RIS is unconfigured and the RIS-based SDR method. Two key observations can be made. Initially, it is evident that the data rate performance remains unaffected by increasing the number of RIS elements in the case of the metal sheet configuration. This outcome aligns with expectations as this approach solely relies on aperture gain without any contribution from passive beamforming gain. Secondly, there is a noticeable contrast in performance between the SDR technique and the metal sheet. The SDR approach consistently outperforms the metal sheet scenario. Notably, the achievable data rate with the SDR approach exhibits a gradual increase in the data rate with the growing number of RIS elements. This progression is logical as the performance of passive beamforming gradually improves.

Figure \ref{fig3} exhibits how the data rate performs across various SNR values, at a fixed 400 RIS elements for the scenarios when the RIS is unconfigured and the RIS-based SDR method. Notably, the superiority of the SDR technique becomes evident compared to the unconfigured surface (metal sheet). This improvement in performance can be attributed to a couple of factors. Firstly, the presence of the RIS enhances the average channel power between the BS and the IoT device. Secondly, the optimization of RIS coefficients contributes to a more constructive integration of the direct and reflected channels at the receiver. As a result, the SDR method exhibits better performance than the unconfigured surface setup. It is also noteworthy that the SDR method outperforms the metal sheet counterpart at both low and high SNR ranges, signifying its superiority across a wide range of signal strengths.

\begin{figure*}[t!]
    \begin{center}    
        \includegraphics[width=1.1\textwidth,trim=20 30 20 0]{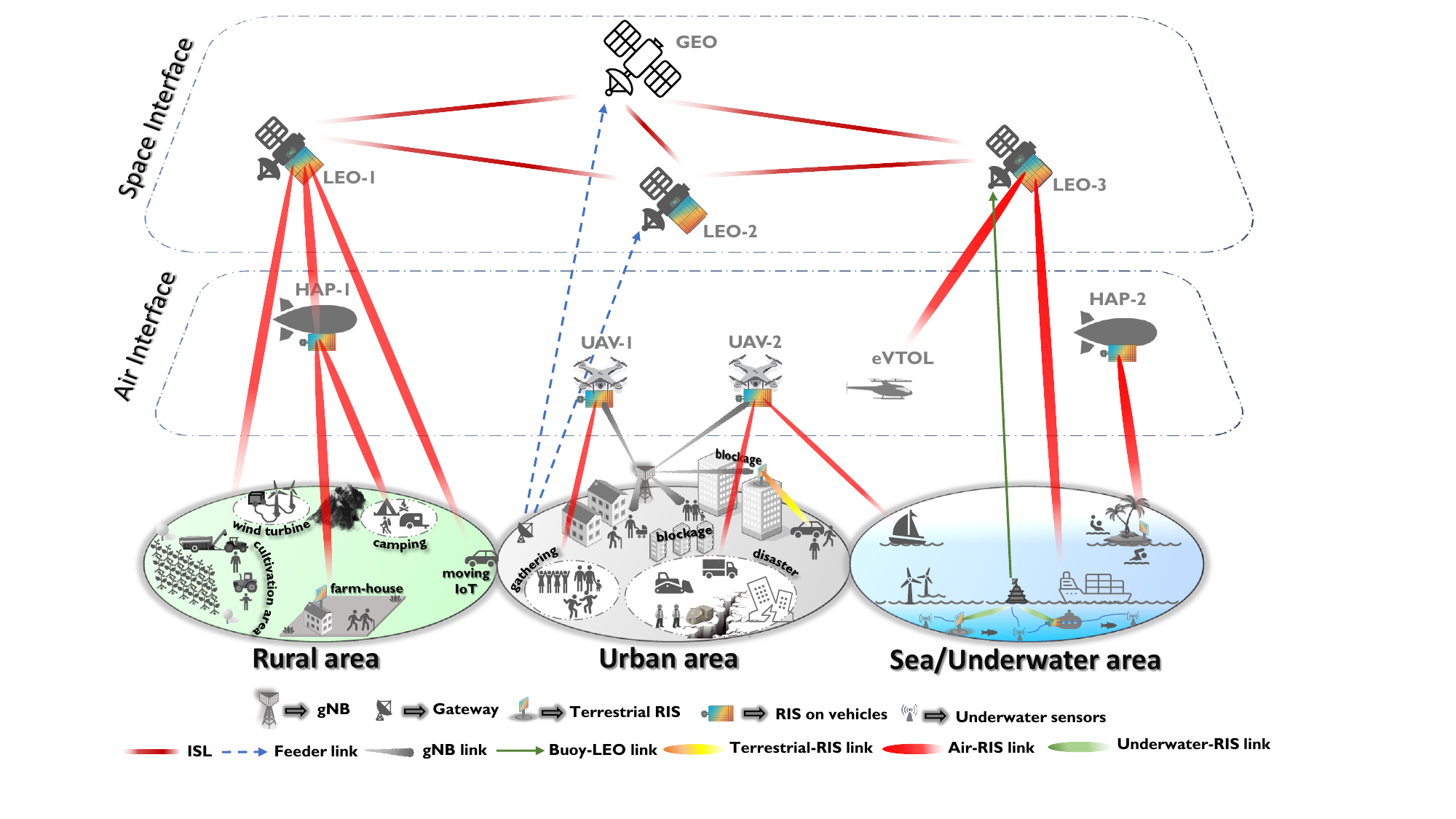}
    \end{center}
    \setlength{\belowcaptionskip}{-2pt}\vspace{-1mm}
    \caption{RIS-aided IoSGT network.}
    \label{IoSGTf}
\end{figure*}

\section{RIS-aided NTN IoT Future Prospect}
\label{sec:5}

So far, the literature has widely presented valuable research on Internet-of-Space-to-Ground-Things (IoSGT), a spatial expansion of IoT. Specifically, IoSGT consists of space networks (including LEO, MEO, and GEO satellites), air networks (including UAV, HAP, and electrically Vertical TakeOff Landing (eVTOL) vehicles), terrestrial networks, maritime/sea networks, and underwater networks. IoSGT supports both horizontal and vertical connectivity with a significant heterogeneity. One step ahead, to ensure ubiquitous 3D connectivity without backhauling issues, it is envisioned that the RIS-aided IoSGT concept will take place in the forthcoming 6G applications \cite{IoSGT}. This is because the existing terrestrial infrastructures lack scalability and limitations in the spectrum resources. Evolving space-to-ground technologies are required to connect massive IoT applications.  

In this regard, several futuristic cases can be considered for rural, urban, and sea/underwater areas, as illustrated in Fig. \ref{IoSGTf}. In rural areas, LEO satellites are the most striking vehicles to serve IoT devices rather than GEO satellites. This is because they have less latency and better channel quality. Thus, demands requested by IoT devices can be fulfilled. On the other side, terrestrial RIS and/or RIS-mounted HAPs can also assist the communication to enhance the link quality and prevent the occurrence of possible dead zones. In terms of an urban area, although terrestrial infrastructures (like gateways) are capable of supporting IoT devices, appropriate cellular coverage may not be possible due to the proper deployment issues and capacity limitations. With GEO satellites, this issue can be overcome at a certain level. Nevertheless, channel conditions can be poor between the GEO and IoT users because of obstructions. In this case, using terrestrial RISs can be a solution; even indoor IoT users can receive a high-quality service via STAR-RISs. Additionally, RIS-mounted LEO satellites and/or UAVs can be used to improve link quality and ensure 3D connectivity. RIS-mounted UAVs are mostly beneficial for urgent incidents and over-capacity usages such as disasters and densely-participant organizations. Further, when an IoT device moves to a rural area, a possible handover issue can be resolved owing to the mega-constellation capability of LEO satellites via ISL links. 

Lastly, many reasons exist to establish IoT networks for sea/underwater areas. There could be disasters such as underwater oil/gas leakage, typhoons, tsunamis, and underwater earthquakes. To monitor physical, chemical, and biological changes underwater is also crucial. Additionally, empowering underwater military applications is important. The most challenging issue in underwater communication is the harsh channel environment. Because the water has a high permittivity and conductivity due to the salinity and temperature, this causes a very high path loss effect. Therefore, acoustic signaling has been proven, and it offers up to tens of km (longer ranges than the electromagnetic wave) with frequencies from Hz to few MHz \cite{UnderwaterRIS}. However, acoustic signaling suffers from scattering at uneven objects, interference of sea animals, and different temperature levels. Hence, it has been envisioned that those imperfections can be mitigated by using RISs. In this regard, RISs can be deployed at the ground of the sea in a stationary manner. On the other hand, they can be mounted on underwater vehicles (like submarines), and also spherical RISs can be deployed at the ground via a cable in a non-stationary manner. In this way, with the aid of RISs, information from sensors can be boosted to buoys and/or ships. Then, the collected information is transmitted to the corresponding departments/institutions via RIS-mounted HAPs and/or LEOs. In terms of serving IoT devices on the sea, RISs play a vital role. Consider islands and many ships on the sea wherein some devices request access to IoT networks. Since they will probably be outside of the coverage of terrestrial infrastructures, RISs can assist them in receiving high-quality information. RIS-mounted LEOs and/or HAPs can serve them with high-quality communication. Also, ships near the sea coast can exploit RIS-mounted UAVs. 

Deploying RISs on structures and airborne/underwater vehicles brings impressive benefits to the IoSGT concept. Nevertheless, to exploit the full benefits of RISs, synchronization, and channel estimation have become a paramount focus. Therefore, the development of robust optimization approaches is required. On top of this, such a heterogeneous connectivity concept (namely IoSGT) requires appropriate interference mitigation techniques without causing overload.  

\section{Conclusion}
The 3GPP-based working groups are making updates to NTN to overcome challenges such as interference, delay, energy efficiency, etc., by spectrum harmonization of satellite and cellular, coverage enhancements, energy savings, etc. In this article, we have demonstrated that integration of RIS with an NTN platform is a reliable solution to fully harness the power of NTN. We have shown that the RIS-aided NTN can provide high spatial resolution, reduce over-the-air signal processing, improve adaptive coverage, and enhance information security. We have also highlighted some of the key challenges associated with RIS-aided UAVs and provided some potential solutions. RIS-assisted UAV deployment scenarios that rely on machine learning techniques to discover coverage holes within the network and expand the network capacity have been discussed. Moreover, we have shown that the RIS-aided UAV network can significantly improve the data rate of the entire communication system in comparison to a non-RIS-based UAV platform. Finally, we have explained the concept of RIS-aided IoSGT network and its significance in integrated terrestrial and NTN architectures.

\bibliographystyle{IEEEtran}

\bibliography{IEEEabrv,BibRef}


\end{document}